\documentclass[%
 reprint,
 amsmath,amssymb,
 aps,
]{revtex4-2}

\usepackage{graphicx}
\usepackage{xcolor}
\usepackage{dcolumn}
\usepackage{bm}
\usepackage{mathtools}

\DeclareMathOperator*{\argmin}{arg\,min}
\DeclarePairedDelimiter{\ceil}{\lceil}{\rceil}

\begin{document}

\preprint{APS/123-QED}

\title{Simulating Large PEPs Tensor Networks on Small Quantum Devices}

\author{Ian MacCormack}
\email{kcamroccamian@gmail.com}
\affiliation{%
 Kadanoff Center for Theoretical Physics, University of Chicago, Chicago, Illinois 60637, USA 
}%
 \affiliation{Department of Physics, Princeton University, Princeton, New Jersey 08544, USA}
\author{Alexey Galda}
\affiliation{James Franck Institute, University of Chicago, Chicago, Illinois 60637, USA}
\affiliation{Computational Science Division, Argonne National Laboratory, Lemont, Illinois 60439, USA}
\affiliation{Menten AI, Inc., Palo Alto, California 94303, USA}

\author{Adam L. Lyon}%
\affiliation{%
Fermi National Accelerator Laboratory, Batavia, Illinois 60510, USA
}%

\date{\today}

\begin{abstract}
We systematically map low-bond-dimension PEPs tensor networks to quantum circuits. By measuring and reusing qubits, we demonstrate that a simulation of an $N \times M$ square-lattice PEPs network, for arbitrary $M$, of bond dimension $2$ can be performed using $N+2$ qubits. We employ this approach to calculate the values of a long-range loop observable in the topological Wen plaquette model by mapping a $3\times 3$ PEPs tensor network to a 5-qubit quantum circuit and executing it on the Honeywell System Model H1-1 trapped-ion device. We find that, for this system size, the noisy observable values are sufficient for diagnosing topological vs. trivial order, as the Wen model is perturbed by a magnetic field term in the Hamiltonian.
Our results serve as a proof-of-concept of the utility of the measure-and-reuse approach for simulating large two-dimensional quantum systems on small quantum devices.

\end{abstract}

\maketitle


\section{Introduction}

As noisy intermediate-scale quantum (NISQ) devices progressively improve in qubit count and gate fidelity, it remains an open question as to whether it is possible to achieve any type of quantum advantage without fully error-corrected fault tolerant algorithms \cite{2021arXiv210108448B}. That is, can near-term quantum processors with $\sim 50-100$ qubits \cite{sarango_earnest-noble_kawase_barkoutsos_2021} perform tasks that a classical computer is either incapable of, or would require significantly more time to perform? A particularly promising avenue for quantum advantage is the simulation of many-body quantum systems~\cite{2016PhRvX...6c1045B}. Indeed, this is a natural niche for quantum computers proposed by Feynman in the early 1980s \cite{feynman1982simulating}.

As the number of qubits offered by modern universal quantum computers remains relatively small, it is advantageous to make greater use of the available quantum resources by measuring and reusing qubits. Such capability has recently become available on some quantum processors~\cite{2021Natur.592..209P}, allowing one to simulate quantum systems consisting of more qubits than are present on the physical device. Several recent works have taken this approach, simulating both static and dynamical one-dimensional (1D) matrix product states (MPS) of quasi-infinite length using a constant number of qubits \cite{2021PhRvR...3c3002F,2021arXiv210411235F,2021arXiv210509324C}. Since most physical quantum states of interest are not maximally entangled, it should not be necessary to use $N$ qubits to simulate most relevant $N$-qubit states. Instead, we propose an adaptation of existing tensor network ans\"atze, which are designed to efficiently represent many-body quantum states on classical computers by exploiting their entanglement structure \cite{2014AnPhy.349..117O}.

Here we take an approach similar to the one used for MPS in \cite{2021PhRvR...3c3002F,2021arXiv210411235F,2021arXiv210509324C}, but with 2D tensor networks. Specifically, we map a subset of projected entangled pair states (PEPs) tensor networks --- which represent 2D quantum states on a square lattice --- to a quantum circuit, and run the resulting circuit on a trapped-ion quantum device. By measuring and reusing qubits, as well as imposing some constraints on the PEPs tensors, we are able to map an $N\times M$ PEPs state to a circuit on $\mathcal{O}(N \log \chi)$ qubits, where $\chi$ is the bond dimension, rather than $\mathcal{O}(N \times M)$ qubits, as the direct mapping would entail. The resulting circuit allows for measurement of arbitrary observables in the PEPs state, without resorting to approximations. Although PEPs tensor networks achieve significant compression, they are not efficiently contractible on classical computers \cite{PhysRevResearch.2.013010}, and their utility as a numerical tool has thus been limited. By outsourcing the contraction problem to a quantum computer, we can exactly probe the physics of larger 2D PEPs states than we could on a classical computer alone.

In this article, we first summarize in Section \ref{the_exp} the experiment we ran on the Honeywell System Model H1-1 trapped-ion device \cite{2021Natur.592..209P,2021arXiv210707505R}, one of a handful of available quantum processors with mid-circuit measure-and-reuse capability. Here, we map a PEPs representation of the topologically-ordered ground state of the Wen plaquette model to a quantum circuit, achieving a compression of $9$ qubits to $5$. We then measure a non-trivial loop observable at various values of a magnetic field perturbation, and compare the results from the ion trap to the classically-contracted tensor network and to exact diagonalization. This allows us to probe the effects of device noise on the computed values of the observable, to establish the feasibility of our approach for larger system sizes. In Section \ref{the_map}, we provide more details about the mapping from PEPs tensor networks to quantum circuits, comparing our approach with that from the recent work \cite{2021arXiv210802792S}. Finally, we close with a discussion of the results and potential future work.

\section{Summary of Experiment}
\label{the_exp}

\begin{figure}[h]
    \centering
    \includegraphics[width=0.5\textwidth]{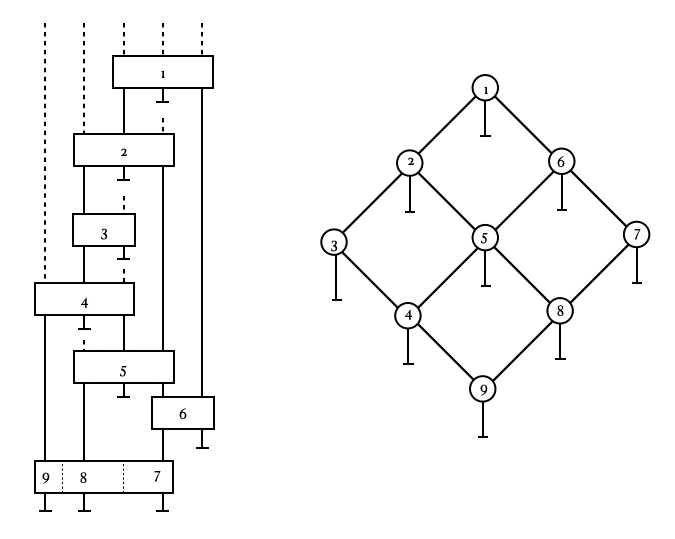}
    \caption{On the left is the 5-qubit circuit corresponding to the 9-qubit PEPs tensor network shown on the right. All qubits are initialized to the zero state at the top of the circuit. The vertical lines truncated by horizontal bars are physical sites and correspond to mid-circuit measurements in the circuit. Dotted lines correspond to qubits in the zero state. After a qubit is measured mid-circuit, it is reset to the zero state.}
    \label{fig:diagram}
\end{figure}

To demonstrate the practicality of our mapping using an existing quantum device, we map a 9-qubit PEPs tensor network with bond dimension $\chi=2$ onto a 5-qubit quantum circuit, with mid-circuit measurements acting as the physical indices of the tensors. A cartoon of the mapping is depicted in Fig. \ref{fig:diagram}.

After optimizing the tensor network (on a classical computer) to reproduce the ground state of a topologically nontrivial Hamiltonian, we compute expectation values of observables in this state by executing the corresponding circuit on the Honeywell System Model H1-1 trapped-ion device. With the resulting data, we then compare the experimental results to exact and tensor network numerical results for these observables.

Our task consists of finding the PEPs representation of the ground state of a particular Hamiltonian, mapping that PEPs to a circuit, and using the circuit to compute non-local order parameters that indicate the phase of the ground state. 

\subsection{Tensor Parameterization and Optimization}

We study the Wen plaquette model \cite{2003PhRvD..68f5003W}:

\begin{equation}
    H_0 = -\sum_{i\in sites} \sigma^x_i \sigma^y_{i+\hat{x}} \sigma^x_{i+\hat{x}+\hat{y}} \sigma^y_{i+\hat{y}},
    \label{ham0}
\end{equation}
where $i+\hat{x}$ and $i+\hat{y}$ indicate neighboring sites to the site $i$ in the $x$ and $y$ directions, respectively. This model's fourfold degenerate ground states (on the plane with open boundary conditions) possess $Z_2$ topological order, which can be diagnosed by measuring a number of different loop operators. Since our 9-site lattice has a boundary, we use the following boundary order parameter (described in \cite{2013PhRvB..87r4402Y}) to detect topological order:

\begin{equation}
    \hat{O}= \sigma^y_1\sigma_2^z\sigma_3^x \sigma_4^z \sigma_9^y \sigma_8^z \sigma_7^x
    \sigma_6^z.
    \label{string}
\end{equation}
This is a closed-loop around the boundary of the square lattice. This operator can be obtained by taking a product of the four plaquette operators in our Hamiltonian. In the perfectly topologically-ordered ground states of (\ref{ham0}), we have $|\langle \hat{O} \rangle | =1$. We introduce the following magnetic field term, as in \cite{2003PhRvD..68f5003W}, to break the topological order:

\begin{equation}
    H_\textrm{mag}= -g \sum_{i\in sites} (\sigma^x_i +\sigma^y_i +\sigma^z_i)
\end{equation}
Our full Hamiltonian is thus $H= H_0+H_\textrm{mag}$. As we increase $g$, the magnetic field strength, the topological order breaks down and $|\langle \hat{O} \rangle | \rightarrow 0$. The exact values of $|\langle \hat{O} \rangle | $ for $g\in [0,1.2]$ are represented by the blue line in Fig. \ref{fig:opplot}.

\begin{figure}
    \centering
    \includegraphics[width=0.5\textwidth]{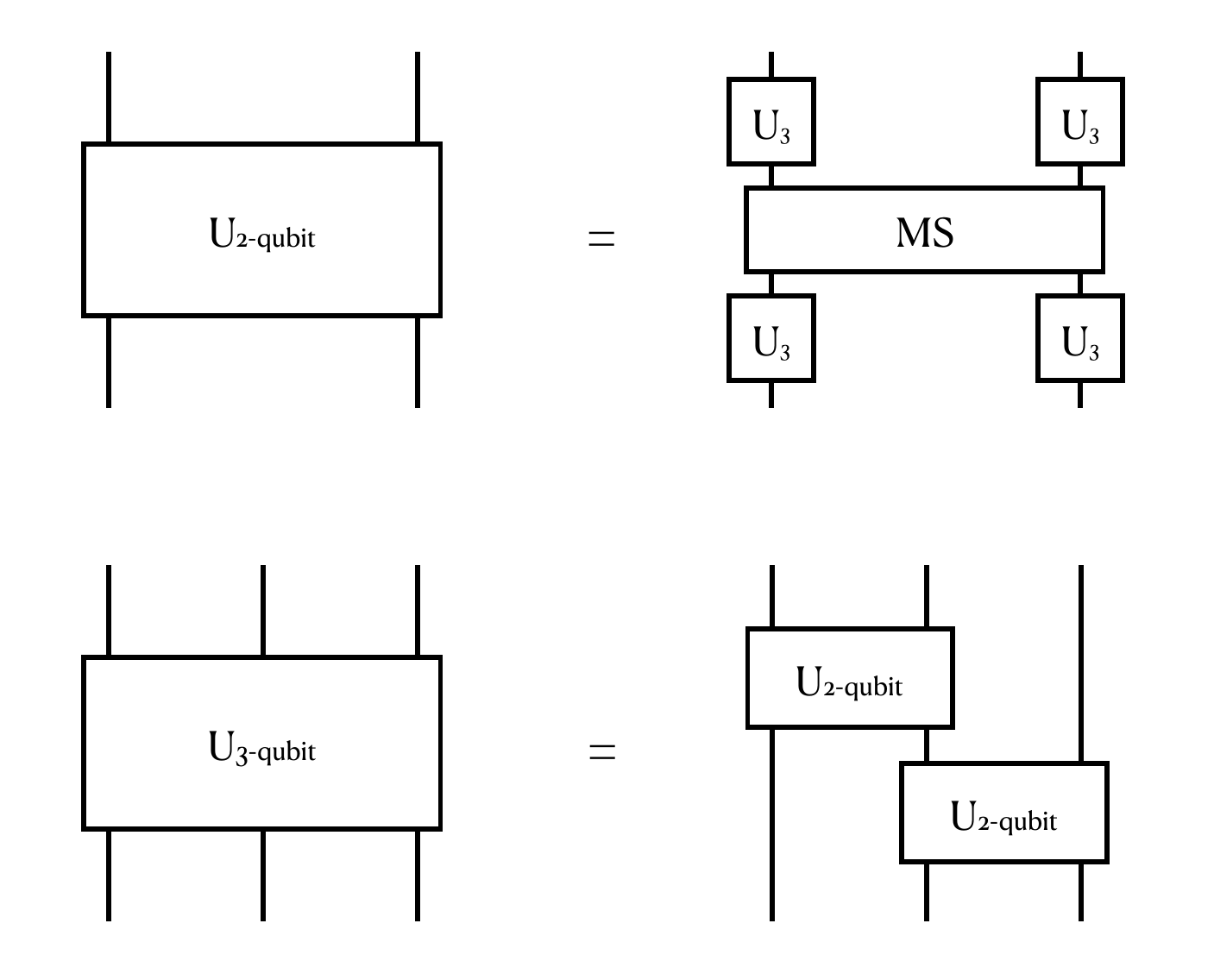}
    \caption{The decomposition of the two and three qubit gates used to parameterize the PEPs circuit. The 2-qubit gate consists of a single M{\o}lmer-S{\o}rensen gate \cite{1999PhRvL..82.1835M} surrounded by an IBM-type $U_3$ rotation \cite{2017arXiv170703429C} on each input and output rail. Each $U_3$ gate contains 3 parameters, giving $U_{2-qubit}$ a total of 12 parameters. Meanwhile, $U_{3-qubit}$, which consists of two, staggered $U_{2-qubit}$ gates, contains 21 independent parameters.}
    \label{fig:decomp}
\end{figure}

With the Hamiltonian in hand, we must first find a PEPs tensor network that approximates the ground state of the system for various values of the magnetic field strength $g$. To parameterize the tensors to be optimized, we start with the unitary circuit elements as depicted on the left-hand side of Fig. \ref{fig:diagram}. The $2$-qubit unitaries in the circuit consist of a single M{\o}lmer-S{\o}rensen (MS) entangling gate \cite{PhysRevLett.82.1835} abutted by a $3$-parameter $SU(2)$ gate on each incoming and outgoing rail (see Fig. \ref{fig:decomp}). Thus, the $2$-qubit unitaries each contain $12$ parameters. The $3$-qubit unitaries in our circuit consist of two staggered $2$-qubit unitaries, and thus contain $24$ parameters (though only $21$ independent parameters). 

The unitary operators are then mapped to tensors by contracting their appropriate auxiliary bonds with the $|0\rangle$ state (depicted by dotted lines in Fig. \ref{fig:diagram}). For the final row, we perform two successive singular value decompositions on the final $3$-qubit unitary to yield the tensors numbered $7$, $8$, and $9$, respectively. The use of a single MS gate in our unitary parameterization ensures that the bond dimension of the resulting tensors is no greater than $2$, as desired. It is worth noting that many of the parameters in the unitary parameterization of the circuit are redundant, as contracting some of the input bonds of these unitaries reduces the effective rank of the resulting map. This is not a major concern when we are optimizing small numbers of parameters ($144$ in this case), but a less redundant parameterization may be necessary for efficient optimization of parameters in larger circuits.

Once we have mapped the parameterized unitaries to tensors, we can contract them into a $3\times3$ square lattice PEPs pattern, as seen on the right panel in Fig.~\ref{fig:diagram}. This is now the parameterized state $|\psi ( \boldsymbol{\theta}) \rangle$, where $\boldsymbol{\theta}$ are the $144$ parameters of the circuit. We then minimize the expectation value of the Hamiltonian to find the tensor network approximation of the ground state for a particular value of $g$ $|GS_\textrm{TN}(g)\rangle$

\begin{equation}
    |GS_\textrm{TN}(g)\rangle= |\psi ( \boldsymbol{\theta}_g) \rangle, \quad \boldsymbol{\theta}_g = \argmin_{\boldsymbol{\theta}} \frac{\langle\psi ( \boldsymbol{\theta}) |H(g) |\psi ( \boldsymbol{\theta}) \rangle}{\langle\psi ( \boldsymbol{\theta}) |\psi ( \boldsymbol{\theta}) \rangle}
\end{equation}
We use the NLopt \cite{Johnson2011} implementation of COBYLA \cite{Powell1994} to perform this optimization.

\subsection{Experimental Results}

\begin{figure*}[ht]
    \centering
    \includegraphics[width=0.7\textwidth]{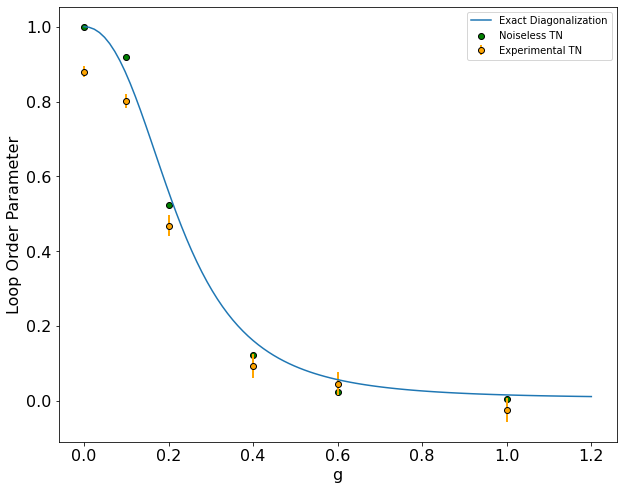}
    \caption{Values of the loop observable (\ref{string}) for various values of the magnetic field strength, $g$. The blue curve indicates the results from exact diagonalization of the Hamiltonian. The green dots are the values of the observable for the tensor network approximation of the ground state (without noise). The orange dots are the experimental results, each corresponding to $1000$ shots of the PEPs circuit on the Honeywell System Model H1-1 device. The error bars on the orange dots correspond to statistical error.}
    \label{fig:opplot}
\end{figure*}

The values of $\langle \hat{O} \rangle $ for the tensor approximations of $|GS(g)\rangle$ for $g=\{0,0.1,0.2,0.4,0.6,1.0 \}$ are depicted as green dots in Fig. \ref{fig:opplot}. These values of the order parameter are not identical to the exact values (blue curve), due to the limited bond dimension of the tensor network. They are, however, close enough to the exact values to distinguish between topological and trivial phases for very high or very low values of $g$.

The effects of the noise from the System Model H1-1 device are evident in the experimental results (orange points). The experiment consisted of $1000$ shots of the PEPs circuit for each separate value of $g$. The error bars on the orange dots are purely from the statistical error from measurements. The error bars are relatively small, so the orange dots provide good data on the effects of noise on the circuit. Overall, the noise has the effect of dampening the values of the loop observable. This effect is more pronounced the larger the expected values of the observable are.
At $g=0$, for example, where $\langle \hat{O} \rangle $ is exactly 1, the results from our circuit yield $\langle \hat{O} \rangle \approx 0.88$, which, absent exact results, serves as a strong indication of topological order. As the strength of the magnetic field $g$ is increased, the measured values of the observable follow the exact tensor network values very closely, and clearly exhibit the transition from topological order to trivial, ferromagnetic order, providing at least a fuzzy phase boundary. The topological order appears to be broken by the time $g$ reaches the value $1$.

The quality of the results for our loop observable indicates that our experiment could potentially by scaled up to larger lattice sizes and still yield useful results on existing quantum hardware.

\section{Mapping Tensor Networks to Quantum Channels}
\label{the_map}

\begin{figure}
    \centering
    \includegraphics[width=0.26\textwidth]{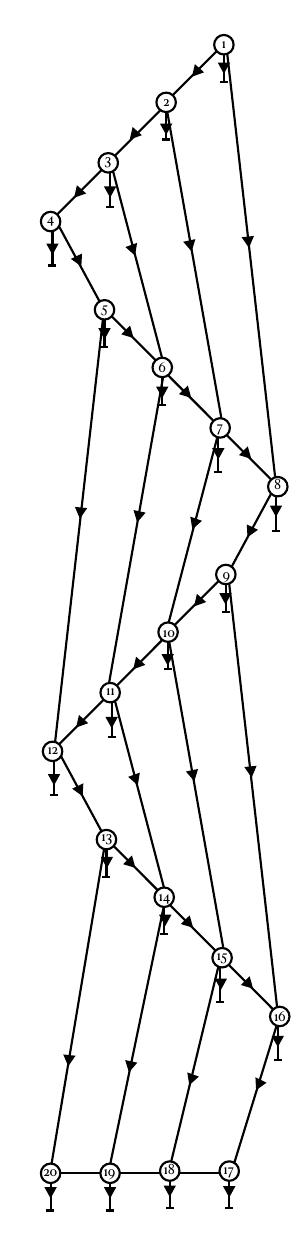}
    \caption{The zig-zag causal pattern used to map a PEPs tensor network onto a quantum circuit, depicted for a $4 \times 5$ qubit rectangular lattice. When mapping to unitary circuit elements, time runs from top to bottom, and physical bonds --- indicated by the dangling lines with orthogonal indicated by dangling lines from tensors with orthogonal crosses --- correspond to mid-circuit measurements followed by qubit resets. }
    \label{fig:zigzag}
\end{figure}

Unlike matrix product states, PEPs tensor networks have no intrinsic ordering that would allow them to be easily mapped to a quantum circuit. Thus, before defining a map from tensors to unitary operators, we must choose an ordering of the contraction of the PEPs tensors that will allow us to construct a quantum circuit. Moreover, we would like to find an ordering that will yield a \textit{qubit-efficient} mapping from PEPs networks to quantum circuits. Ideally, we want to make use of mid-circuit measurement and reset so that our circuit contains fewer qubits than are present in the corresponding PEPs state, along the lines of the ``holographic" simulation of MPS tensor networks employed in \cite{2021PhRvR...3c3002F,2021arXiv210411235F,2021arXiv210509324C}. To this end, we order the PEPs tensor network in a zig-zag pattern, as depicted in Fig. \ref{fig:zigzag}. The incoming and outgoing arrows on the physical and virtual bonds of a tensor indicate the flow of time, and correspond to the input and output indices of a unitary circuit element, respectively. 

\begin{figure}
    \centering
    \includegraphics[width=0.35\textwidth]{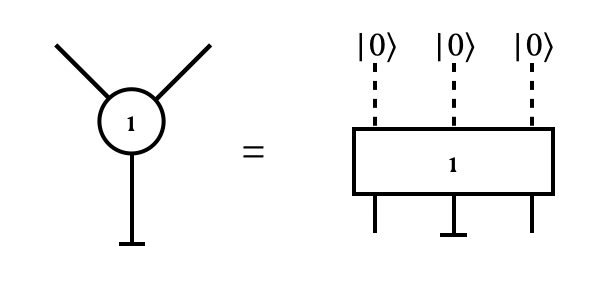}
    \includegraphics[width=0.3\textwidth]{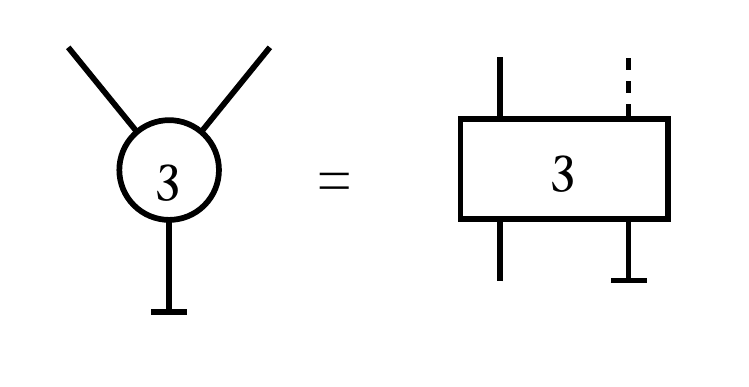}
    \caption{A mapping of tensors $1$ and $3$, respectively, from the causally-ordered tensor network depicted in Fig. \ref{fig:diagram} to unitary operators containing auxiliary input legs, which are always initialized to the zero state. For the unitary operators on the right and side, time flows from top to bottom. }
    \label{fig:tensor_ops}
\end{figure}

Tensors that have more output than input indices can be supplemented with additional input indices (which will be fixed to the $|0\rangle$ state) in order to embed them in a unitary matrix. For example, the tensor numbered $1$ in Fig. \ref{fig:zigzag} contains three output and zero input bonds. Two of the output bonds are virtual bonds, each of bond dimension $\chi$ and will thus require $N_\textrm{B}= \ceil{\log \chi }$ qubits as virtual qubits in a corresponding unitary. Thus, a corresponding unitary operator is a $2N_\textrm{B}+1$ qubit unitary with all of its input qubits initialized in the $|0\rangle$ state. Similarly, tensor $3$ will be mapped to a unitary operator on $N_\textrm{B}+1$ qubits, with one of the input qubits set to $|0\rangle$ --- $N_\textrm{B}$ of the qubits in the unitary correspond to the virtual bonds, while the additional qubit corresponds to the physical bond. Diagrammatic examples of this supplementation of bonds can be seen in Fig. \ref{fig:tensor_ops}. It is important to note that all supplementary bonds are input bonds. This allows the input qubits to these bonds to be deterministically set to $|0\rangle$ before the execution of the unitary gate. This allows us to avoid post-selection, an issue that would arise with output bonds that must be set to $|0\rangle$.

With this zig-zag causal ordering in place, we can map PEPs tensor networks on rectangular lattices of size $N \times M$ for arbitrary $M$ with bond dimension $\chi=2^{N_\textrm{B}}$ to quantum circuits on $N_\textrm{Q}=(N+1)\times N_\textrm{B} +1$ qubits. This becomes qubit-efficient, that is $N_\textrm{Q}< N\times M$, when

\begin{equation}
    M > \left \lfloor \left( 1 +\frac{1}{N} \right) N_\textrm{B} + \frac{1}{N} \right \rfloor,
\end{equation}

or, in terms of the number of bond qubits,

\begin{equation}
    N_\textrm{B} <\left \lfloor \frac{N\times M -1}{N+1}\right \rfloor.
\end{equation}

For example, for a $3\times3$ square lattice, we must have a bond dimension of $2$ or lower for our circuit to be qubit-efficient. For a general $N\times N$ square lattice, the circuit is qubit-efficient if $N_\textrm{B} \leq (N-1)$. Thus, for a given bond dimension, the circuit mapping becomes more qubit-efficient as we increase the size of the lattice.

\subsection{Parameterizing the PEPs Tensors}
\label{mapping}

One can choose a variety of tensor decomposition methods as a means to parameterize the PEPs tensor network, and, ultimately, its corresponding quantum channel. In parameterizing tensor $3$ in Figs. \ref{fig:diagram} and \ref{fig:tensor_ops}, for example, starting with an arbitrary rank-$3$ tensor (assuming $\chi =2$ for now), $T_i^{\alpha \beta}$ (where Roman letters indicate physical bonds and Greek letters indicate virtual bonds and we use Einstein summation), one can construct a $2$-qubit unitary $U_{i \beta, \alpha \gamma}$ such that when one of the input qubits is set to the zero state (as depicted in Fig. \ref{fig:tensor_ops}), we have

\begin{equation}
    U_{i \beta, \alpha \gamma} \delta_{\gamma,0}=U_{i \beta, \alpha 0} = T_i^{\alpha \beta}.
\end{equation}

Once this constraint is in place, we must ensure the unitarity of $U_{i \beta, \alpha \gamma}$:

\begin{gather}
    U^*_{\lambda \sigma,i \beta} U_{i \beta, \alpha \gamma} = \delta_{\lambda \sigma,\alpha \gamma} \\
    U^*_{\lambda 0,i \beta} U_{i \beta, \alpha 0}=T_i^{* \lambda \beta} T_i^{ \beta \alpha}
    = \delta_{\lambda ,\alpha }.
\end{gather}

This first constraint indicates that the tensors must be isometric in one direction (that is, when we contract a bond index and a physical index of two tensors, we obtain the identity for the remaining bond indices). Expanding the second constraint,


\begin{align}
    U_{j \zeta, \alpha \gamma} U^*_{\alpha \gamma, i \beta} &= \delta_{j \zeta,i \beta} \notag\\\notag
    &= U_{j \zeta, \alpha 0} U^*_{\alpha 0, i \beta} + U_{j \zeta, \alpha 1} U^*_{\alpha 1, i \beta}\\
    &= T_j^{ \zeta \alpha} T_i^{* \alpha \beta} + U_{j \zeta, \alpha 1} U^*_{\alpha 1, i \beta}
     \delta_{j \zeta,i \beta}\\
    U_{j \zeta, \alpha 1} U^*_{\alpha 1, i \beta}
    &=\delta_{j \zeta,i \beta} -T_j^{ \zeta \alpha} T_i^{* \alpha \beta}.
\end{align}

This second constraint allows us to solve for the remaining components of the unitary matrix using the components of the tensor.

\subsection{Comparison with Isometric Tensor Networks}
\label{isotns}

The constraints we impose on the PEPs tensors in order to define a causal structure are related to those proposed for so-called ``isometric tensor networks" (isoTNS) \cite{2020PhRvL.124c7201Z}. In that work, the authors outline a procedure that involves first choosing a site (the ``orthogonality center") in a 2D tensor network on which one wishes to compute the expectation value of a local observable. The column and row that intersect at this site then become the ``orthogonality hypersurface". The tensors constituting the complementary regions of this hypersurface are then chosen to be isometric, with orthogonality to the hypersurfaces defining the ``causal order" along which tensor contractions are isometric. In our zig-zag causal structure, no orthogonality center is specified, and the direction of isometry is different than that implied by the isoTNS method if we had chosen the final row of tensors as an orthogonality hypersurface. In our causal order, alternating rows have alternating causal order, whereas in the isoTNS structure, rows of tensors parallel to the orthogonality hypersurface have unidirectional causal order.

The disadvantage of the isoTNS approach is that one must select an orthogonality center before computing an observable, making it inconvenient for computing large or non-local observables, such as Wilson loops. In a recent work by Slattery and Clark \cite{2021arXiv210802792S}, the authors propose ``quantum isometric tensor networks" (qisoTNS), which are very similar, but not identical to our proposal. In that work, as in ours, one does not need to restrict to a selected orthogonality hypersurface in order to compute an observable. The work by Slattery and Clark retains the causal structure implied by isoTNS, unlike our zig-zag causal structure, but achieves similar compression of $N\times N$ square lattices to $\mathcal{O}(N \log \chi)$ qubit circuits, as our work does.

\subsection{Mapping the Final Row of Tensors}

For the final row of tensors evaluated in the circuit (i.e. tensors $17, 18, 19, $ and $20$ in Fig. \ref{fig:zigzag}), one may either continue the zig-zag pattern employed for the previous rows, or use one large $N$-qubit unitary operator to contain all $N$ tensors in the row. In the former case, the mapping from tensors to unitaries proceeds as outlined in \ref{mapping}. In the latter case, one can simply contract the $N$ tensors of the final row together to form an $N$-qubit operator, and constrain the tensors such that this operator is unitary. This is what we have done to parameterize the tensor network in our experiment, as it slightly reduces the depth of the circuit as compared to the case where we retain zig-zag ordering in the final row.

\section{Conclusion and Outlook}

In this work we have confirmed the possibility of obtaining useful results from a PEPs to quantum circuit mapping on existing quantum hardware, at least for small system sizes. The obvious next step is to scale up the system size and determine the point at which device noise washes out the values of the desired observables and the results are no longer useful as e.g. probes of topological order. The H1-1 device supports up to 10 qubits, allowing for PEPs tensor networks with bond dimension 2 of up to $8\times M$ (for arbitrary M) qubits to be simulated. The utility of the results and their resilience to noise will vary depending on the system and the measured observables, but measuring a long range loop order parameter in a topological phase, as we have done in this paper, could serve as a fairly sensitive benchmark for the susceptibility of the circuit to noise.

Several other extensions of this work could also prove to be valuable. As suggested in \cite{2021arXiv210802792S}, it is not exactly clear how expressive PEPs tensor networks with isometric constraints can be. Clearly, the causal structure imposed by mapping PEPs to a circuit will limit what correlations can and cannot be present in the system. It may also prevent the exact imposition of certain spatial symmetries on the system. Whether this limitation can be overcome by simply increasing the bond dimension, or if this sort of PEPs tensor network is limited to a certain subset of 2D states is an interesting open question. One advantage of this construction is the direct access it allows to the bonds of the PEPs tensor network, which should allow for fairly straightforward computation of various entanglement measures. It would be very interesting, via direct tomography or some other method, to use this approach to experimentally measure, e.g. the topological entanglement entropy in a 2D state \cite{2006PhRvL..96k0405L,2006PhRvL..96k0404K}. Additionally, it may be interesting to construct an analogous mapping for non-square lattice tensor network states. More broadly, it is worth asking if and how one could achieve some sort of quantum advantage using an approach like this. This would likely involve finding a noise-resilient observable in a specifically-chosen interacting 2D model, and measuring it using a quantum circuit of a size that cannot be efficiently classically simulated. Attempting to construct such a problem would be an interesting and worthwhile project. Regardless, it is likely that NISQ devices will at least serve as important companions to classical simulation in the study of strongly-correlated many-body quantum systems in the very near future.

\begin{acknowledgments}

We thank Matthew Otten and Shinsei Ryu for useful discussions and feedback. We would also like to thank Brian Neyenhuis, James Goeders and the rest of the Honeywell QC team for assistance with their device. IM is supported through Shinsei Ryu by a Simons Investigator Grant from the Simons Foundation. This work was supported in part by the National Science Foundation grant DMR 2001181. We thank Fermilab for providing access to the Honeywell System Model H1-1 under laboratory directed research and development project FNAL-LDRD-2018-025. AL is supported by Fermi Research Alliance, LLC under Contract No. DE-AC02-07CH11359 with the U.S. Department of Energy (DOE), Office of Science, Office of High Energy Physics (OHEP) and by DOE OHEP QuantISED program grant: Large Scale Simulations of Quantum Systems on HPC with Analytics for HEP Algorithms (0000246788).

\end{acknowledgments}

\nocite{*}

\bibliography{apssamp}

\begin{thebibliography}{22}%
\makeatletter
\providecommand \@ifxundefined [1]{%
 \@ifx{#1\undefined}
}%
\providecommand \@ifnum [1]{%
 \ifnum #1\expandafter \@firstoftwo
 \else \expandafter \@secondoftwo
 \fi
}%
\providecommand \@ifx [1]{%
 \ifx #1\expandafter \@firstoftwo
 \else \expandafter \@secondoftwo
 \fi
}%
\providecommand \natexlab [1]{#1}%
\providecommand \enquote  [1]{``#1''}%
\providecommand \bibnamefont  [1]{#1}%
\providecommand \bibfnamefont [1]{#1}%
\providecommand \citenamefont [1]{#1}%
\providecommand \href@noop [0]{\@secondoftwo}%
\providecommand \href [0]{\begingroup \@sanitize@url \@href}%
\providecommand \@href[1]{\@@startlink{#1}\@@href}%
\providecommand \@@href[1]{\endgroup#1\@@endlink}%
\providecommand \@sanitize@url [0]{\catcode `\\12\catcode `\$12\catcode
  `\&12\catcode `\#12\catcode `\^12\catcode `\_12\catcode `\%12\relax}%
\providecommand \@@startlink[1]{}%
\providecommand \@@endlink[0]{}%
\providecommand \url  [0]{\begingroup\@sanitize@url \@url }%
\providecommand \@url [1]{\endgroup\@href {#1}{\urlprefix }}%
\providecommand \urlprefix  [0]{URL }%
\providecommand \Eprint [0]{\href }%
\providecommand \doibase [0]{https://doi.org/}%
\providecommand \selectlanguage [0]{\@gobble}%
\providecommand \bibinfo  [0]{\@secondoftwo}%
\providecommand \bibfield  [0]{\@secondoftwo}%
\providecommand \translation [1]{[#1]}%
\providecommand \BibitemOpen [0]{}%
\providecommand \bibitemStop [0]{}%
\providecommand \bibitemNoStop [0]{.\EOS\space}%
\providecommand \EOS [0]{\spacefactor3000\relax}%
\providecommand \BibitemShut  [1]{\csname bibitem#1\endcsname}%
\let\auto@bib@innerbib\@empty
\bibitem [{\citenamefont {{Bharti}}\ \emph {et~al.}(2021)\citenamefont
  {{Bharti}}, \citenamefont {{Cervera-Lierta}}, \citenamefont {{Kyaw}},
  \citenamefont {{Haug}}, \citenamefont {{Alperin-Lea}}, \citenamefont
  {{Anand}}, \citenamefont {{Degroote}}, \citenamefont {{Heimonen}},
  \citenamefont {{Kottmann}}, \citenamefont {{Menke}}, \citenamefont {{Mok}},
  \citenamefont {{Sim}}, \citenamefont {{Kwek}},\ and\ \citenamefont
  {{Aspuru-Guzik}}}]{2021arXiv210108448B}%
  \BibitemOpen
  \bibfield  {author} {\bibinfo {author} {\bibfnamefont {K.}~\bibnamefont
  {{Bharti}}}, \bibinfo {author} {\bibfnamefont {A.}~\bibnamefont
  {{Cervera-Lierta}}}, \bibinfo {author} {\bibfnamefont {T.~H.}\ \bibnamefont
  {{Kyaw}}}, \bibinfo {author} {\bibfnamefont {T.}~\bibnamefont {{Haug}}},
  \bibinfo {author} {\bibfnamefont {S.}~\bibnamefont {{Alperin-Lea}}}, \bibinfo
  {author} {\bibfnamefont {A.}~\bibnamefont {{Anand}}}, \bibinfo {author}
  {\bibfnamefont {M.}~\bibnamefont {{Degroote}}}, \bibinfo {author}
  {\bibfnamefont {H.}~\bibnamefont {{Heimonen}}}, \bibinfo {author}
  {\bibfnamefont {J.~S.}\ \bibnamefont {{Kottmann}}}, \bibinfo {author}
  {\bibfnamefont {T.}~\bibnamefont {{Menke}}}, \bibinfo {author} {\bibfnamefont
  {W.-K.}\ \bibnamefont {{Mok}}}, \bibinfo {author} {\bibfnamefont
  {S.}~\bibnamefont {{Sim}}}, \bibinfo {author} {\bibfnamefont {L.-C.}\
  \bibnamefont {{Kwek}}},\ and\ \bibinfo {author} {\bibfnamefont
  {A.}~\bibnamefont {{Aspuru-Guzik}}},\ }\bibfield  {title} {\bibinfo {title}
  {{Noisy intermediate-scale quantum (NISQ) algorithms}},\ }\href@noop {}
  {\bibfield  {journal} {\bibinfo  {journal} {arXiv e-prints}\ ,\ \bibinfo
  {eid} {arXiv:2101.08448}} (\bibinfo {year} {2021})},\ \Eprint
  {https://arxiv.org/abs/2101.08448} {arXiv:2101.08448 [quant-ph]} \BibitemShut
  {NoStop}%
\bibitem [{\citenamefont {Sarango}\ \emph {et~al.}(2021)\citenamefont
  {Sarango}, \citenamefont {Saul}, \citenamefont {Earnest-Noble}, \citenamefont
  {Nate}, \citenamefont {Kawase}, \citenamefont {Kei}, \citenamefont
  {Barkoutsos},\ and\ \citenamefont
  {Panagiotis}}]{sarango_earnest-noble_kawase_barkoutsos_2021}%
  \BibitemOpen
  \bibfield  {author} {\bibinfo {author} {\bibfnamefont {R.~Z.}\ \bibnamefont
  {Sarango}}, \bibinfo {author} {\bibnamefont {Saul}}, \bibinfo {author}
  {\bibfnamefont {S.~E.}\ \bibnamefont {Earnest-Noble}}, \bibinfo {author}
  {\bibnamefont {Nate}}, \bibinfo {author} {\bibfnamefont {J.~G.}\ \bibnamefont
  {Kawase}}, \bibinfo {author} {\bibnamefont {Kei}}, \bibinfo {author}
  {\bibfnamefont {I.~T.}\ \bibnamefont {Barkoutsos}},\ and\ \bibinfo {author}
  {\bibnamefont {Panagiotis}},\ }\href
  {https://research.ibm.com/blog/ibm-quantum-roadmap} {\bibinfo {title} {Ibm's
  roadmap for scaling quantum technology}} (\bibinfo {year} {2021})\BibitemShut
  {NoStop}%
\bibitem [{\citenamefont {{Bauer}}\ \emph {et~al.}(2016)\citenamefont
  {{Bauer}}, \citenamefont {{Wecker}}, \citenamefont {{Millis}}, \citenamefont
  {{Hastings}},\ and\ \citenamefont {{Troyer}}}]{2016PhRvX...6c1045B}%
  \BibitemOpen
  \bibfield  {author} {\bibinfo {author} {\bibfnamefont {B.}~\bibnamefont
  {{Bauer}}}, \bibinfo {author} {\bibfnamefont {D.}~\bibnamefont {{Wecker}}},
  \bibinfo {author} {\bibfnamefont {A.~J.}\ \bibnamefont {{Millis}}}, \bibinfo
  {author} {\bibfnamefont {M.~B.}\ \bibnamefont {{Hastings}}},\ and\ \bibinfo
  {author} {\bibfnamefont {M.}~\bibnamefont {{Troyer}}},\ }\bibfield  {title}
  {\bibinfo {title} {{Hybrid Quantum-Classical Approach to Correlated
  Materials}},\ }\href {https://doi.org/10.1103/PhysRevX.6.031045} {\bibfield
  {journal} {\bibinfo  {journal} {Physical Review X}\ }\textbf {\bibinfo
  {volume} {6}},\ \bibinfo {eid} {031045} (\bibinfo {year} {2016})},\ \Eprint
  {https://arxiv.org/abs/1510.03859} {arXiv:1510.03859 [quant-ph]} \BibitemShut
  {NoStop}%
\bibitem [{\citenamefont {Feynman}(1982)}]{feynman1982simulating}%
  \BibitemOpen
  \bibfield  {author} {\bibinfo {author} {\bibfnamefont {R.~P.}\ \bibnamefont
  {Feynman}},\ }\bibfield  {title} {\bibinfo {title} {Simulating physics with
  computers},\ }\href@noop {} {\bibfield  {journal} {\bibinfo  {journal}
  {International journal of theoretical physics}\ }\textbf {\bibinfo {volume}
  {21}},\ \bibinfo {pages} {467} (\bibinfo {year} {1982})}\BibitemShut
  {NoStop}%
\bibitem [{\citenamefont {{Pino}}\ \emph {et~al.}(2021)\citenamefont {{Pino}},
  \citenamefont {{Dreiling}}, \citenamefont {{Figgatt}}, \citenamefont
  {{Gaebler}}, \citenamefont {{Moses}}, \citenamefont {{Allman}}, \citenamefont
  {{Baldwin}}, \citenamefont {{Foss-Feig}}, \citenamefont {{Hayes}},
  \citenamefont {{Mayer}}, \citenamefont {{Ryan-Anderson}},\ and\ \citenamefont
  {{Neyenhuis}}}]{2021Natur.592..209P}%
  \BibitemOpen
  \bibfield  {author} {\bibinfo {author} {\bibfnamefont {J.~M.}\ \bibnamefont
  {{Pino}}}, \bibinfo {author} {\bibfnamefont {J.~M.}\ \bibnamefont
  {{Dreiling}}}, \bibinfo {author} {\bibfnamefont {C.}~\bibnamefont
  {{Figgatt}}}, \bibinfo {author} {\bibfnamefont {J.~P.}\ \bibnamefont
  {{Gaebler}}}, \bibinfo {author} {\bibfnamefont {S.~A.}\ \bibnamefont
  {{Moses}}}, \bibinfo {author} {\bibfnamefont {M.~S.}\ \bibnamefont
  {{Allman}}}, \bibinfo {author} {\bibfnamefont {C.~H.}\ \bibnamefont
  {{Baldwin}}}, \bibinfo {author} {\bibfnamefont {M.}~\bibnamefont
  {{Foss-Feig}}}, \bibinfo {author} {\bibfnamefont {D.}~\bibnamefont
  {{Hayes}}}, \bibinfo {author} {\bibfnamefont {K.}~\bibnamefont {{Mayer}}},
  \bibinfo {author} {\bibfnamefont {C.}~\bibnamefont {{Ryan-Anderson}}},\ and\
  \bibinfo {author} {\bibfnamefont {B.}~\bibnamefont {{Neyenhuis}}},\
  }\bibfield  {title} {\bibinfo {title} {{Demonstration of the trapped-ion
  quantum CCD computer architecture}},\ }\href
  {https://doi.org/10.1038/s41586-021-03318-4} {\bibfield  {journal} {\bibinfo
  {journal} {\nat}\ }\textbf {\bibinfo {volume} {592}},\ \bibinfo {pages} {209}
  (\bibinfo {year} {2021})},\ \Eprint {https://arxiv.org/abs/2003.01293}
  {arXiv:2003.01293 [quant-ph]} \BibitemShut {NoStop}%
\bibitem [{\citenamefont {{Foss-Feig}}\ \emph
  {et~al.}(2021{\natexlab{a}})\citenamefont {{Foss-Feig}}, \citenamefont
  {{Hayes}}, \citenamefont {{Dreiling}}, \citenamefont {{Figgatt}},
  \citenamefont {{Gaebler}}, \citenamefont {{Moses}}, \citenamefont {{Pino}},\
  and\ \citenamefont {{Potter}}}]{2021PhRvR...3c3002F}%
  \BibitemOpen
  \bibfield  {author} {\bibinfo {author} {\bibfnamefont {M.}~\bibnamefont
  {{Foss-Feig}}}, \bibinfo {author} {\bibfnamefont {D.}~\bibnamefont
  {{Hayes}}}, \bibinfo {author} {\bibfnamefont {J.~M.}\ \bibnamefont
  {{Dreiling}}}, \bibinfo {author} {\bibfnamefont {C.}~\bibnamefont
  {{Figgatt}}}, \bibinfo {author} {\bibfnamefont {J.~P.}\ \bibnamefont
  {{Gaebler}}}, \bibinfo {author} {\bibfnamefont {S.~A.}\ \bibnamefont
  {{Moses}}}, \bibinfo {author} {\bibfnamefont {J.~M.}\ \bibnamefont
  {{Pino}}},\ and\ \bibinfo {author} {\bibfnamefont {A.~C.}\ \bibnamefont
  {{Potter}}},\ }\bibfield  {title} {\bibinfo {title} {{Holographic quantum
  algorithms for simulating correlated spin systems}},\ }\href
  {https://doi.org/10.1103/PhysRevResearch.3.033002} {\bibfield  {journal}
  {\bibinfo  {journal} {Physical Review Research}\ }\textbf {\bibinfo {volume}
  {3}},\ \bibinfo {eid} {033002} (\bibinfo {year} {2021}{\natexlab{a}})},\
  \Eprint {https://arxiv.org/abs/2005.03023} {arXiv:2005.03023 [quant-ph]}
  \BibitemShut {NoStop}%
\bibitem [{\citenamefont {{Foss-Feig}}\ \emph
  {et~al.}(2021{\natexlab{b}})\citenamefont {{Foss-Feig}}, \citenamefont
  {{Ragole}}, \citenamefont {{Potter}}, \citenamefont {{Dreiling}},
  \citenamefont {{Figgatt}}, \citenamefont {{Gaebler}}, \citenamefont {{Hall}},
  \citenamefont {{Moses}}, \citenamefont {{Pino}}, \citenamefont {{Spaun}},
  \citenamefont {{Neyenhuis}},\ and\ \citenamefont
  {{Hayes}}}]{2021arXiv210411235F}%
  \BibitemOpen
  \bibfield  {author} {\bibinfo {author} {\bibfnamefont {M.}~\bibnamefont
  {{Foss-Feig}}}, \bibinfo {author} {\bibfnamefont {S.}~\bibnamefont
  {{Ragole}}}, \bibinfo {author} {\bibfnamefont {A.}~\bibnamefont {{Potter}}},
  \bibinfo {author} {\bibfnamefont {J.}~\bibnamefont {{Dreiling}}}, \bibinfo
  {author} {\bibfnamefont {C.}~\bibnamefont {{Figgatt}}}, \bibinfo {author}
  {\bibfnamefont {J.}~\bibnamefont {{Gaebler}}}, \bibinfo {author}
  {\bibfnamefont {A.}~\bibnamefont {{Hall}}}, \bibinfo {author} {\bibfnamefont
  {S.}~\bibnamefont {{Moses}}}, \bibinfo {author} {\bibfnamefont
  {J.}~\bibnamefont {{Pino}}}, \bibinfo {author} {\bibfnamefont
  {B.}~\bibnamefont {{Spaun}}}, \bibinfo {author} {\bibfnamefont
  {B.}~\bibnamefont {{Neyenhuis}}},\ and\ \bibinfo {author} {\bibfnamefont
  {D.}~\bibnamefont {{Hayes}}},\ }\bibfield  {title} {\bibinfo {title}
  {{Entanglement from tensor networks on a trapped-ion QCCD quantum
  computer}},\ }\href@noop {} {\bibfield  {journal} {\bibinfo  {journal} {arXiv
  e-prints}\ ,\ \bibinfo {eid} {arXiv:2104.11235}} (\bibinfo {year}
  {2021}{\natexlab{b}})},\ \Eprint {https://arxiv.org/abs/2104.11235}
  {arXiv:2104.11235 [quant-ph]} \BibitemShut {NoStop}%
\bibitem [{\citenamefont {{Chertkov}}\ \emph {et~al.}(2021)\citenamefont
  {{Chertkov}}, \citenamefont {{Bohnet}}, \citenamefont {{Francois}},
  \citenamefont {{Gaebler}}, \citenamefont {{Gresh}}, \citenamefont {{Hankin}},
  \citenamefont {{Lee}}, \citenamefont {{Tobey}}, \citenamefont {{Hayes}},
  \citenamefont {{Neyenhuis}}, \citenamefont {{Stutz}}, \citenamefont
  {{Potter}},\ and\ \citenamefont {{Foss-Feig}}}]{2021arXiv210509324C}%
  \BibitemOpen
  \bibfield  {author} {\bibinfo {author} {\bibfnamefont {E.}~\bibnamefont
  {{Chertkov}}}, \bibinfo {author} {\bibfnamefont {J.}~\bibnamefont
  {{Bohnet}}}, \bibinfo {author} {\bibfnamefont {D.}~\bibnamefont
  {{Francois}}}, \bibinfo {author} {\bibfnamefont {J.}~\bibnamefont
  {{Gaebler}}}, \bibinfo {author} {\bibfnamefont {D.}~\bibnamefont {{Gresh}}},
  \bibinfo {author} {\bibfnamefont {A.}~\bibnamefont {{Hankin}}}, \bibinfo
  {author} {\bibfnamefont {K.}~\bibnamefont {{Lee}}}, \bibinfo {author}
  {\bibfnamefont {R.}~\bibnamefont {{Tobey}}}, \bibinfo {author} {\bibfnamefont
  {D.}~\bibnamefont {{Hayes}}}, \bibinfo {author} {\bibfnamefont
  {B.}~\bibnamefont {{Neyenhuis}}}, \bibinfo {author} {\bibfnamefont
  {R.}~\bibnamefont {{Stutz}}}, \bibinfo {author} {\bibfnamefont {A.~C.}\
  \bibnamefont {{Potter}}},\ and\ \bibinfo {author} {\bibfnamefont
  {M.}~\bibnamefont {{Foss-Feig}}},\ }\bibfield  {title} {\bibinfo {title}
  {{Holographic dynamics simulations with a trapped ion quantum computer}},\
  }\href@noop {} {\bibfield  {journal} {\bibinfo  {journal} {arXiv e-prints}\
  ,\ \bibinfo {eid} {arXiv:2105.09324}} (\bibinfo {year} {2021})},\ \Eprint
  {https://arxiv.org/abs/2105.09324} {arXiv:2105.09324 [quant-ph]} \BibitemShut
  {NoStop}%
\bibitem [{\citenamefont {{Or{\'u}s}}(2014)}]{2014AnPhy.349..117O}%
  \BibitemOpen
  \bibfield  {author} {\bibinfo {author} {\bibfnamefont {R.}~\bibnamefont
  {{Or{\'u}s}}},\ }\bibfield  {title} {\bibinfo {title} {{A practical
  introduction to tensor networks: Matrix product states and projected
  entangled pair states}},\ }\href {https://doi.org/10.1016/j.aop.2014.06.013}
  {\bibfield  {journal} {\bibinfo  {journal} {Annals of Physics}\ }\textbf
  {\bibinfo {volume} {349}},\ \bibinfo {pages} {117} (\bibinfo {year}
  {2014})},\ \Eprint {https://arxiv.org/abs/1306.2164} {arXiv:1306.2164
  [cond-mat.str-el]} \BibitemShut {NoStop}%
\bibitem [{\citenamefont {Haferkamp}\ \emph {et~al.}(2020)\citenamefont
  {Haferkamp}, \citenamefont {Hangleiter}, \citenamefont {Eisert},\ and\
  \citenamefont {Gluza}}]{PhysRevResearch.2.013010}%
  \BibitemOpen
  \bibfield  {author} {\bibinfo {author} {\bibfnamefont {J.}~\bibnamefont
  {Haferkamp}}, \bibinfo {author} {\bibfnamefont {D.}~\bibnamefont
  {Hangleiter}}, \bibinfo {author} {\bibfnamefont {J.}~\bibnamefont {Eisert}},\
  and\ \bibinfo {author} {\bibfnamefont {M.}~\bibnamefont {Gluza}},\ }\bibfield
   {title} {\bibinfo {title} {Contracting projected entangled pair states is
  average-case hard},\ }\href
  {https://doi.org/10.1103/PhysRevResearch.2.013010} {\bibfield  {journal}
  {\bibinfo  {journal} {Phys. Rev. Research}\ }\textbf {\bibinfo {volume}
  {2}},\ \bibinfo {pages} {013010} (\bibinfo {year} {2020})}\BibitemShut
  {NoStop}%
\bibitem [{\citenamefont {{Ryan-Anderson}}\ \emph {et~al.}(2021)\citenamefont
  {{Ryan-Anderson}}, \citenamefont {{Bohnet}}, \citenamefont {{Lee}},
  \citenamefont {{Gresh}}, \citenamefont {{Hankin}}, \citenamefont {{Gaebler}},
  \citenamefont {{Francois}}, \citenamefont {{Chernoguzov}}, \citenamefont
  {{Lucchetti}}, \citenamefont {{Brown}}, \citenamefont {{Gatterman}},
  \citenamefont {{Halit}}, \citenamefont {{Gilmore}}, \citenamefont {{Gerber}},
  \citenamefont {{Neyenhuis}}, \citenamefont {{Hayes}},\ and\ \citenamefont
  {{Stutz}}}]{2021arXiv210707505R}%
  \BibitemOpen
  \bibfield  {author} {\bibinfo {author} {\bibfnamefont {C.}~\bibnamefont
  {{Ryan-Anderson}}}, \bibinfo {author} {\bibfnamefont {J.~G.}\ \bibnamefont
  {{Bohnet}}}, \bibinfo {author} {\bibfnamefont {K.}~\bibnamefont {{Lee}}},
  \bibinfo {author} {\bibfnamefont {D.}~\bibnamefont {{Gresh}}}, \bibinfo
  {author} {\bibfnamefont {A.}~\bibnamefont {{Hankin}}}, \bibinfo {author}
  {\bibfnamefont {J.~P.}\ \bibnamefont {{Gaebler}}}, \bibinfo {author}
  {\bibfnamefont {D.}~\bibnamefont {{Francois}}}, \bibinfo {author}
  {\bibfnamefont {A.}~\bibnamefont {{Chernoguzov}}}, \bibinfo {author}
  {\bibfnamefont {D.}~\bibnamefont {{Lucchetti}}}, \bibinfo {author}
  {\bibfnamefont {N.~C.}\ \bibnamefont {{Brown}}}, \bibinfo {author}
  {\bibfnamefont {T.~M.}\ \bibnamefont {{Gatterman}}}, \bibinfo {author}
  {\bibfnamefont {S.~K.}\ \bibnamefont {{Halit}}}, \bibinfo {author}
  {\bibfnamefont {K.}~\bibnamefont {{Gilmore}}}, \bibinfo {author}
  {\bibfnamefont {J.}~\bibnamefont {{Gerber}}}, \bibinfo {author}
  {\bibfnamefont {B.}~\bibnamefont {{Neyenhuis}}}, \bibinfo {author}
  {\bibfnamefont {D.}~\bibnamefont {{Hayes}}},\ and\ \bibinfo {author}
  {\bibfnamefont {R.~P.}\ \bibnamefont {{Stutz}}},\ }\bibfield  {title}
  {\bibinfo {title} {{Realization of real-time fault-tolerant quantum error
  correction}},\ }\href@noop {} {\bibfield  {journal} {\bibinfo  {journal}
  {arXiv e-prints}\ ,\ \bibinfo {eid} {arXiv:2107.07505}} (\bibinfo {year}
  {2021})},\ \Eprint {https://arxiv.org/abs/2107.07505} {arXiv:2107.07505
  [quant-ph]} \BibitemShut {NoStop}%
\bibitem [{\citenamefont {{Slattery}}\ and\ \citenamefont
  {{Clark}}(2021)}]{2021arXiv210802792S}%
  \BibitemOpen
  \bibfield  {author} {\bibinfo {author} {\bibfnamefont {L.}~\bibnamefont
  {{Slattery}}}\ and\ \bibinfo {author} {\bibfnamefont {B.~K.}\ \bibnamefont
  {{Clark}}},\ }\bibfield  {title} {\bibinfo {title} {{Quantum Circuits For
  Two-Dimensional Isometric Tensor Networks}},\ }\href@noop {} {\bibfield
  {journal} {\bibinfo  {journal} {arXiv e-prints}\ ,\ \bibinfo {eid}
  {arXiv:2108.02792}} (\bibinfo {year} {2021})},\ \Eprint
  {https://arxiv.org/abs/2108.02792} {arXiv:2108.02792 [quant-ph]} \BibitemShut
  {NoStop}%
\bibitem [{\citenamefont {{Wen}}(2003)}]{2003PhRvD..68f5003W}%
  \BibitemOpen
  \bibfield  {author} {\bibinfo {author} {\bibfnamefont {X.-G.}\ \bibnamefont
  {{Wen}}},\ }\bibfield  {title} {\bibinfo {title} {{Quantum order from
  string-net condensations and the origin of light and massless fermions}},\
  }\href {https://doi.org/10.1103/PhysRevD.68.065003} {\bibfield  {journal}
  {\bibinfo  {journal} {\prd}\ }\textbf {\bibinfo {volume} {68}},\ \bibinfo
  {eid} {065003} (\bibinfo {year} {2003})},\ \Eprint
  {https://arxiv.org/abs/hep-th/0302201} {arXiv:hep-th/0302201 [hep-th]}
  \BibitemShut {NoStop}%
\bibitem [{\citenamefont {{Yu}}\ \emph {et~al.}(2013)\citenamefont {{Yu}},
  \citenamefont {{Zhang}},\ and\ \citenamefont {{Kou}}}]{2013PhRvB..87r4402Y}%
  \BibitemOpen
  \bibfield  {author} {\bibinfo {author} {\bibfnamefont {J.}~\bibnamefont
  {{Yu}}}, \bibinfo {author} {\bibfnamefont {X.-H.}\ \bibnamefont {{Zhang}}},\
  and\ \bibinfo {author} {\bibfnamefont {S.-P.}\ \bibnamefont {{Kou}}},\
  }\bibfield  {title} {\bibinfo {title} {{Majorana edge states for Z$_{2}$
  topological orders of the Wen plaquette and toric code models}},\ }\href
  {https://doi.org/10.1103/PhysRevB.87.184402} {\bibfield  {journal} {\bibinfo
  {journal} {\prb}\ }\textbf {\bibinfo {volume} {87}},\ \bibinfo {eid} {184402}
  (\bibinfo {year} {2013})},\ \Eprint {https://arxiv.org/abs/1209.5460}
  {arXiv:1209.5460 [cond-mat.str-el]} \BibitemShut {NoStop}%
\bibitem [{\citenamefont {M\o{}lmer}\ and\ \citenamefont
  {S\o{}rensen}(1999)}]{PhysRevLett.82.1835}%
  \BibitemOpen
  \bibfield  {author} {\bibinfo {author} {\bibfnamefont {K.}~\bibnamefont
  {M\o{}lmer}}\ and\ \bibinfo {author} {\bibfnamefont {A.}~\bibnamefont
  {S\o{}rensen}},\ }\bibfield  {title} {\bibinfo {title} {Multiparticle
  entanglement of hot trapped ions},\ }\href
  {https://doi.org/10.1103/PhysRevLett.82.1835} {\bibfield  {journal} {\bibinfo
   {journal} {Phys. Rev. Lett.}\ }\textbf {\bibinfo {volume} {82}},\ \bibinfo
  {pages} {1835} (\bibinfo {year} {1999})}\BibitemShut {NoStop}%
\bibitem [{\citenamefont {{M{\o}lmer}}\ and\ \citenamefont
  {{S{\o}rensen}}(1999)}]{1999PhRvL..82.1835M}%
  \BibitemOpen
  \bibfield  {author} {\bibinfo {author} {\bibfnamefont {K.}~\bibnamefont
  {{M{\o}lmer}}}\ and\ \bibinfo {author} {\bibfnamefont {A.}~\bibnamefont
  {{S{\o}rensen}}},\ }\bibfield  {title} {\bibinfo {title} {{Multiparticle
  Entanglement of Hot Trapped Ions}},\ }\href
  {https://doi.org/10.1103/PhysRevLett.82.1835} {\bibfield  {journal} {\bibinfo
   {journal} {\prl}\ }\textbf {\bibinfo {volume} {82}},\ \bibinfo {pages}
  {1835} (\bibinfo {year} {1999})},\ \Eprint
  {https://arxiv.org/abs/quant-ph/9810040} {arXiv:quant-ph/9810040 [quant-ph]}
  \BibitemShut {NoStop}%
\bibitem [{\citenamefont {{Cross}}\ \emph {et~al.}(2017)\citenamefont
  {{Cross}}, \citenamefont {{Bishop}}, \citenamefont {{Smolin}},\ and\
  \citenamefont {{Gambetta}}}]{2017arXiv170703429C}%
  \BibitemOpen
  \bibfield  {author} {\bibinfo {author} {\bibfnamefont {A.~W.}\ \bibnamefont
  {{Cross}}}, \bibinfo {author} {\bibfnamefont {L.~S.}\ \bibnamefont
  {{Bishop}}}, \bibinfo {author} {\bibfnamefont {J.~A.}\ \bibnamefont
  {{Smolin}}},\ and\ \bibinfo {author} {\bibfnamefont {J.~M.}\ \bibnamefont
  {{Gambetta}}},\ }\bibfield  {title} {\bibinfo {title} {{Open Quantum Assembly
  Language}},\ }\href@noop {} {\bibfield  {journal} {\bibinfo  {journal} {arXiv
  e-prints}\ ,\ \bibinfo {eid} {arXiv:1707.03429}} (\bibinfo {year} {2017})},\
  \Eprint {https://arxiv.org/abs/1707.03429} {arXiv:1707.03429 [quant-ph]}
  \BibitemShut {NoStop}%
\bibitem [{\citenamefont {Johnson}(2011)}]{Johnson2011}%
  \BibitemOpen
  \bibfield  {author} {\bibinfo {author} {\bibfnamefont {S.~G.}\ \bibnamefont
  {Johnson}},\ }\href {http://ab-initio.mit.edu/nlopt} {\emph {\bibinfo {title}
  {The NLopt nonlinear-optimization package}}} (\bibinfo {year}
  {2011})\BibitemShut {NoStop}%
\bibitem [{\citenamefont {Powell}(1994)}]{Powell1994}%
  \BibitemOpen
  \bibfield  {author} {\bibinfo {author} {\bibfnamefont {M.~J.~D.}\
  \bibnamefont {Powell}},\ }\bibinfo {title} {A direct search optimization
  method that models the objective and constraint functions by linear
  interpolation},\ in\ \href {https://doi.org/10.1007/978-94-015-8330-5_4}
  {\emph {\bibinfo {booktitle} {Advances in Optimization and Numerical
  Analysis}}},\ \bibinfo {editor} {edited by\ \bibinfo {editor} {\bibfnamefont
  {S.}~\bibnamefont {Gomez}}\ and\ \bibinfo {editor} {\bibfnamefont {J.-P.}\
  \bibnamefont {Hennart}}}\ (\bibinfo  {publisher} {Springer Netherlands},\
  \bibinfo {address} {Dordrecht},\ \bibinfo {year} {1994})\ pp.\ \bibinfo
  {pages} {51--67}\BibitemShut {NoStop}%
\bibitem [{\citenamefont {{Zaletel}}\ and\ \citenamefont
  {{Pollmann}}(2020)}]{2020PhRvL.124c7201Z}%
  \BibitemOpen
  \bibfield  {author} {\bibinfo {author} {\bibfnamefont {M.~P.}\ \bibnamefont
  {{Zaletel}}}\ and\ \bibinfo {author} {\bibfnamefont {F.}~\bibnamefont
  {{Pollmann}}},\ }\bibfield  {title} {\bibinfo {title} {{Isometric Tensor
  Network States in Two Dimensions}},\ }\href
  {https://doi.org/10.1103/PhysRevLett.124.037201} {\bibfield  {journal}
  {\bibinfo  {journal} {\prl}\ }\textbf {\bibinfo {volume} {124}},\ \bibinfo
  {eid} {037201} (\bibinfo {year} {2020})},\ \Eprint
  {https://arxiv.org/abs/1902.05100} {arXiv:1902.05100 [cond-mat.str-el]}
  \BibitemShut {NoStop}%
\bibitem [{\citenamefont {{Levin}}\ and\ \citenamefont
  {{Wen}}(2006)}]{2006PhRvL..96k0405L}%
  \BibitemOpen
  \bibfield  {author} {\bibinfo {author} {\bibfnamefont {M.}~\bibnamefont
  {{Levin}}}\ and\ \bibinfo {author} {\bibfnamefont {X.-G.}\ \bibnamefont
  {{Wen}}},\ }\bibfield  {title} {\bibinfo {title} {{Detecting Topological
  Order in a Ground State Wave Function}},\ }\href
  {https://doi.org/10.1103/PhysRevLett.96.110405} {\bibfield  {journal}
  {\bibinfo  {journal} {\prl}\ }\textbf {\bibinfo {volume} {96}},\ \bibinfo
  {eid} {110405} (\bibinfo {year} {2006})},\ \Eprint
  {https://arxiv.org/abs/cond-mat/0510613} {arXiv:cond-mat/0510613
  [cond-mat.str-el]} \BibitemShut {NoStop}%
\bibitem [{\citenamefont {{Kitaev}}\ and\ \citenamefont
  {{Preskill}}(2006)}]{2006PhRvL..96k0404K}%
  \BibitemOpen
  \bibfield  {author} {\bibinfo {author} {\bibfnamefont {A.}~\bibnamefont
  {{Kitaev}}}\ and\ \bibinfo {author} {\bibfnamefont {J.}~\bibnamefont
  {{Preskill}}},\ }\bibfield  {title} {\bibinfo {title} {{Topological
  Entanglement Entropy}},\ }\href
  {https://doi.org/10.1103/PhysRevLett.96.110404} {\bibfield  {journal}
  {\bibinfo  {journal} {\prl}\ }\textbf {\bibinfo {volume} {96}},\ \bibinfo
  {eid} {110404} (\bibinfo {year} {2006})},\ \Eprint
  {https://arxiv.org/abs/hep-th/0510092} {arXiv:hep-th/0510092 [hep-th]}
  \BibitemShut {NoStop}%
\end{thebibliography}%

\end{document}